\documentclass[twocolumn,aps,pra,showpacs,floatfix,superscriptaddress]{revtex4}

\usepackage{epsf}
\usepackage{amsmath}
\usepackage{amsfonts}
\usepackage{amssymb}
\usepackage{bm}
\usepackage{bbm}
\usepackage{nicefrac}
\usepackage{color}
\usepackage{pifont}
\usepackage{graphicx}
\usepackage{enumerate}
\usepackage{dcolumn}
\usepackage{maybemath}

\newcolumntype{.}{D{x}{}{6}}

\def\alphaQED{\alpha_{\rm QED}}

\def\dd{{\mathrm{d}}}
\def\ii{{\mathrm{i}}}

\def\calE{{\mathcal{E}}}
\def\calO{{\mathcal{O}}}
\def\calV{{\mathcal{V}}}

\def\calZ{{\mathcal{Z}}}

\def\tfrac#1#2{ {\textstyle{\frac{#1}{#2}} } }

\begin{document}

\title{Functional Form of the Imaginary Part of the Atomic Polarizability}

\author{U. D. Jentschura}

\affiliation{Department of Physics, Missouri University of Science
and Technology, Rolla, Missouri 65409, USA}

\author{K. Pachucki}

\affiliation{Faculty of Physics, University of Warsaw,
Pasteura 5, 02--093 Warsaw, Poland}

\begin{abstract}
The dynamic atomic polarizability describes the response of the atom
to incoming electromagnetic radiation.  The functional form of the imaginary
part of the polarizability for small driving frequencies $\omega$ has been a
matter of long-standing discussion, with both a linear dependence and an
$\omega^3$ dependence being presented as candidate formulas. The imaginary part
of the polarizability enters the expressions of a number of fundamental
physical processes which involve the thermal dissipation of energy, such as
blackbody friction, and non-contact friction.  Here, we solve the long-standing
problem by calculating the imaginary part of the polarizability 
in both the length (``$\vec d \cdot \vec E$'') as
well as the velocity-gauge (``$\vec p \cdot \vec A$'') form of the dipole
interaction, verify the gauge invariance,
and find general expressions applicable to atomic theory; the
$\omega^3$ form is obtained in both gauges. The seagull term in the velocity
gauge is found to be crucial in establishing gauge invariance. 
\end{abstract} 


\maketitle


%
%
\section{Introduction}
\label{sec1}

The ``susceptibility'' of an atom toward the generation of an atomic dipole
moment is described by the dynamic atomic (dipole) polarizability.
Alternatively, the atomic polarizability describes
photon absorption from a field of photons and
subsequent emission of a photon into the same or a vacuum mode
of different wave vector but the same frequency.
The external oscillating
field can be a laser field (and this is the intuitive 
picture we shall use in this paper).
In this case, the real part of the
polarizability describes the ac Stark shift~\cite{HaJeKe2006}.
The atomic polarizability has both a real as well as an 
imaginary part. Physically, this can be illustrated by 
considering the relative permittivity $\epsilon_r(\omega)$ 
of a dilute gas and its relation to the dynamic
dipole polarizability $\alpha(\omega)$ of the gas atoms,
\begin{equation}
\label{one}
\epsilon_r(\omega) = 1 + \frac{N_V}{\epsilon_0} \alpha(\omega) \,,
\end{equation}
where $\epsilon_0$ is the vacuum permittivity 
and $N_V$ is the volume density of gas atoms~\cite{Ja1998}.
The Kramers--Kronig relations dictate that 
the real part of $\epsilon(\omega) $ should be an even 
function of the driving frequency $\omega$,
while the imaginary part of $\epsilon(\omega)$ cannot vanish and 
should be an odd function of $\omega$.
These considerations are valid upon an interpretation 
of the dielectric constant in terms of the retarded 
Green function $G_R$ which describes the relation of the 
dielectric displacement $\vec D(\vec r, t)$ to the 
electric field $\vec E(\vec r, t)$, 
\begin{equation}
\vec D(\vec r, t) = 
\epsilon_0 \, \vec E(\vec r, t) + 
\epsilon_0 \int_0^\infty \dd \tau \, 
G_R(\tau) \, \vec E(\vec r, t - \tau) \,.
\end{equation}
The Fourier transform is 
\begin{equation}
G_R(\omega) = \frac{N_V}{\epsilon_0} \alpha(\omega) \,.
\end{equation}
The symmetry properties of the imaginary part of the 
polarizability have been discussed at various 
places in the literature~\cite{AnDRSt2003,MiBo2004,MiLoBeBa2008,%
WaEtAl2009,InEtAl2011}. The conventions used here 
and also the result of our calculation
of the non-resonant contribution to the imaginary part, 
to be reported in the following, 
are consistent with the interpretation as a 
retarded Green function and, notably, with the conclusions of 
Ref.~\cite{MiBo2004} [see Eq.~(2) and the text following 
Eq.~(31) of Ref.~\cite{MiBo2004}].

The imaginary part of the polarizability enters the description 
of processes such as the black-body friction~\cite{MkPaPoSa2004,LaDKJe2012prl},
where an atom is decelerated by interaction with a thermal bath of 
black-body photons. The blackbody (BB) friction force 
$F_{\rm BB} = -\eta_{\rm BB} \, v$ is linear in the velocity,
\begin{equation}
\label{force1}
\eta_{\rm BB} = 
\frac{\beta\hbar^2}{12\pi^2\,\epsilon_0 \, c^5} \,
\int\limits_0^{\infty}
\frac{\dd\omega\,\omega^5\; {\rm Im}[\alpha(\omega)]}
{\sinh^2(\tfrac12 \beta \hbar \omega)} \,.
\end{equation}
Here, $\beta = 1/(k_B \, T)$ is the Boltzmann factor,
$c$ is the speed of light, and $\hbar$ is the 
quantum unit of action (reduced Planck constant),
while the integration is carried out over the entire
blackbody spectrum of angular frequencies $\omega$.
This force is due to dissipative processes; 
the atom absorbs an incoming blue-shifted 
photon from the front, and emits photons in 
all directions.
The friction force at a distance $\calZ$ 
above a surface composed of a dielectric 
material, due to the dragging of the 
mirror charges inside the electric with 
frequency-dependent dielectric constant 
$\epsilon(\omega)$, is calculated as $F_{\rm QF} = -\eta_{\rm QF} \, v$.
According to Ref.~\cite{ToWi1997}, the result
for the non-contact quantum friction (QF) 
coefficient in the relation is given by a Green--Kubo formula,
\begin{align}
\label{force2}
\eta_{\rm QF} = & \; 
\frac{3 \beta \hbar^2}{32 \pi^2 \epsilon_0  \calZ^5} \, \int\limits_{0}^\infty 
\frac{\dd \omega \,
{\rm Im} [\alpha(\omega)]}{\sinh^2(\tfrac12 \beta \hbar \omega)} 
{\rm Im} \left[ \frac{\epsilon(\omega)-1}{\epsilon(\omega)+1} \right] \,.
\end{align}

The formulas~\eqref{force1} and~\eqref{force2} raise the question what
could be deemed to be the most consistent physical picture behind the imaginary
part of the polarizability.  
The imaginary part of the polarizability involves a spontaneous photon emission
process, and this spontaneous emission can only be understood if one quantizes
the electromagnetic field. From a quantum theory point of view, the reference
state in the calculation of the leading-order polarizability is given by the
atom in the reference ground state and $n_L$ photons in the laser field.  The
virtual states for the calculation of the polarizability involve the atom in a
virtual excited state, and $n_L \pm 1$ photons in the laser field.  The dipole
coupling leads to the creation or annihilation of a virtual laser photon.

Let us try to calculate the imaginary part of the polarizability on the basis
of an energy shift calculation.  In many cases, the imaginary part of an energy
shift describes a decay process, with the spontaneous emission of radiation
quanta.  If one includes the energy of the emitted quanta into the energy
balance, one sees that the final quantum state of the decay process actually
has the same energy as the initial state.  We consider as an example the
spontaneous emission of a photon by an atom. The initial quantum state has the
atom in an excited state, and zero photons in the radiation field.  In the
final state of the process, the atom is in the ground state, and we have one
photon in the radiation field, with an energy that exactly compensates the
energy loss of the atom.  The imaginary part of the self-energy of an excited
state describes the decay width of the excited reference state, against
one-photon decay~\cite{BaSu1978}.  This consideration has recently been
generalized to the two-photon self-energy~\cite{Je2007}, which helped clarify
the role of cascade contributions which need to be separated from the coherent
two-photon correction to the decay rate~\cite{Je2010}.

How can this intuitive picture be applied to the imaginary part of the
polarizability? From a quantum theory point of view~\cite{HaJeKe2006}, the
initial state/reference state in the calculation of the polarizability involves
the atom in the ground state, and $n_L$ photons in the radiation field.  The
laser photons are not in resonance with any atomic transition, but rather, in a
typical calculation, of very low frequency.  One may ask what is the
energetically degenerate state to which the reference state could ``decay''.
In the case of the ground-state atomic polarizability, one may additionally
point out that the atom already is in the ground state and cannot go
energetically lower. How could the ``decay rate'' be formulated under these
conditions?  The answer is that a quantum state with the atom in the ground
state, $n_L - 1$ laser photons of energy $\omega_L$, and one photon of wave
vector $\vec k$ and polarization $\lambda$, with energy $\omega_{\vec k
\lambda} = \omega_L$ (but not the same polarization or propagation direction)
is energetically degenerate with respect to the reference state (with the atom
still in the ground state, and $n_L$ laser photons).  The reference state in
our calculation will thus be the product state of atom in the reference state
(in general the ground state) $|\phi \rangle$, the laser field in the state
with $n_L$ photons, and zero photons in other modes of the quantized
electromagnetic field.  We denote this state as
\begin{equation}
|\phi_0\rangle = | \phi, n_L, 0 \rangle \,.
\end{equation}
The calculation of the energy shift then proceeds in analogy to
the self-energy calculation: One inserts radiative loops 
into the diagrams that describe the ac Stark shift and 
calculates the decay to the state
\begin{equation}
|\phi_f\rangle = | \phi, n_L-1, 1_{\vec k \, \lambda} \rangle \,.
\end{equation}
The notation adopted here involves the occupation numbers
of the laser mode (subscript $L$), and the 
mode with wave vector $\vec k$ and polarization $\lambda$.

The remainder of this paper is organized as follows.  In Sec.~\ref{sec2}, we
discuss the derivation of the imaginary part of the polarizability in the
length gauge, where the interaction of the atom with the laser field, and also
the interaction of the atom with the quantized radiation field, are modeled on
the basis of the dipole interaction $-e \, \vec r\cdot \vec E$.  Gauge
invariance (see Sec.~\ref{sec3}) is used as a method to verify our results.  In
the velocity gauge, we use the ``dipole'' coupling $-e \, \vec p\cdot \vec A/m$
and the ``seagull term'' $e^2 \vec A^2/(2 m)$. Here, $\vec A$ is the vector
potential, while $\vec E = -\partial_t \vec A$ is the electric field.  We
employ the dipole approximation throughout the paper and work in SI mksA units
(with the exception of a few illustrative unit conversions in Sec.~\ref{sec4}).

%
%
\section{Length Gauge}
\label{sec2}

We start with the length-gauge calculation.
The laser is $z$ polarized, and we consider the electric 
dipole interaction which is the dominant interaction for an 
atom. The electric field operators for the 
laser field (subscript $L$) and for the 
quantized field interaction (subscript $I$) read as follows,
\begin{subequations}
\label{HLI}
\begin{align}
\vec E_L = & \; \hat e_z \, \sqrt{\frac{\hbar \omega_L}{2 \epsilon_0 \calV_L}} \,
\left( a_L + a_L^+ \right) 
= \hat e_z \, E_L \,, 
\\[0.133ex]
\vec E_I =& \; 
\sum_{\vec k \, \lambda} 
\sqrt{\frac{\hbar \omega_{\vec k \, \lambda}}{2 \epsilon_0 \calV}} \,
\hat\epsilon_{\vec k\, \lambda}
\left( a_{\vec k \, \lambda} + a_{\vec k \, \lambda}^+ \right) \,,
\\[0.133ex]
H_L =& \; -e \, z \, E_L  \,,
\qquad 
H_I = -e \, \vec r \cdot \vec E_I \,.
\end{align}
\end{subequations}
We employ the dipole approximation, setting 
the spatial phase factors $\exp(\ii \vec k \cdot \vec r)$ 
equal to unity.
The use of an explicit normalization volume 
$\calV_L$ for the laser mode (and $\calV$ for the modes of the 
quantized field) allows for a consistent normalization 
of the laser intensity [see Eq.~\eqref{IL} below].
The sum over the modes of the vacuum in 
$H_I$ explicitly excludes the laser mode, 
in the sense of the requirement
$\vec k \, \lambda \neq L$. However, the absence of the 
one mode does not enter the matching condition
\begin{equation}
\label{match}
\sum_{\vec k} =
\calV \, \int \frac{\dd^3 k}{(2 \pi)^3} \,,
\end{equation}
because the one highly occupied laser mode 
does not feature prominently in the sum over all modes,
and the sum over polarizations can be carried out using the 
relation
\begin{equation}
\label{sumpol}
\sum_\lambda \hat\epsilon^i_{\vec k\, \lambda} \,
\hat\epsilon^j_{\vec k\, \lambda} = 
\delta^{ij} - \frac{k^i\,k^j}{\vec k^{\,2}} \,.
\end{equation}
The quantity $\calV_L$ is the normalization volume for the 
laser field, whereas $\calV$ is the corresponding 
term for the quantized field mode.
The laser field intensity is 
\begin{equation}
\label{IL}
I_L = \frac{n_L \hbar \omega_L \, c}{\calV_L} \,.
\end{equation}
The unperturbed Hamiltonian is given as follows,
\begin{subequations}
\begin{align}
H_0 =& \; H_A + H_{EM} \,,
\\[0.133ex]
H_A =& \;
\sum_m E_m \, | \phi_m \rangle \, \langle \phi_m | \,,
\\[0.133ex]
H_{EM} =& \; 
\sum_{\vec k \, \lambda \neq L} \hbar \omega_{\vec k \, \lambda}
a_{\vec k \, \lambda}^+ \; a_{\vec k \, \lambda} +
\hbar \omega_L \; a^+_L \; a_L \,.
\end{align}
\end{subequations}
Here, the subscript $A$ stands for atom,
while the subscript $EM$ denotes the 
(quantized) electromagnetic field.
The unperturbed state is given by
\begin{equation}
| \phi_0 \rangle = | \phi, \, n_L, 0 \rangle  \,,
\qquad
H_A \, | \phi \rangle = E \, |\phi \rangle \,.
\end{equation}
denoting the atom in the reference state $| \phi \rangle$
(notably, the ground state), 
$n_L$ photons in the laser mode, and the vacuum 
state for the remaining modes of the electromagnetic field.
We should add that the Fock state of the laser 
field is just a calculational device for our 
derivation of the imaginary part of the polarizability.
(Of course, in a Fock state, 
the occupation number is precisely $n_L$,
which is in contrast a coherent state which is a superposition of Fock states.)

The atomic Hamiltonian (Schr\"{o}dinger Hamiltonian) 
is denoted as $H_A$, 
and we do all calculations for hydrogen, here,
whereas for many-electron atoms, in the nonrelativistic 
approximation, one would replace $z \to \sum z_a$, 
where $a$ denotes the $a$th electron. 
All results below will be reported for atomic hydrogen 
with a single electron coordinate, but the results 
hold more generally, with the only necessary 
modification for a many-electron atom being the 
addition of the coordinates of the other electrons.
The reference state fulfills the relations
\begin{equation}
H_0 \, | \phi_0 \rangle = E_0  \, | \phi_0 \rangle  \,,
\qquad
E_0 = E + n_L \hbar \omega_L \,,
\end{equation}
where $n_L$ is the number of laser photons.

In the quantized formalism~\cite{HaJeKe2006}, the 
second-order energy shift, which we use for normalization,
can be calculated as
\begin{equation}
\delta E^{(2)} =
-\left< H_L \, \frac{1}{(H_0 - E_0)'}  \, H_L \right> \,,
\end{equation}
where the prime denotes the reduced Green function.
(Replacement of $H_L$ by $H_I$ leads to the
low-energy part of the Lamb shift, see Ref.~\cite{JeKe2004aop}.
Here, the Lamb shift is absorbed into a renormalized reference
state energy $E$.)
In the matrix elements used in this paper,
the reference state is the full quantized-field 
reference state $|\phi_0 \rangle$ unless
indicated differently.
After tracing out the photon degrees of freedom,
one obtains
\begin{align}
\delta E^{(2)} =& \;
-e^2 \, \frac{n_L \hbar \omega_L}{2 \epsilon_0 \calV_L}
\left(
\left< \phi \left| z \, 
\left( \frac{1}{H_A - E + \hbar \omega_L} \right) \, z 
\right.  \right.  \right.
\nonumber\\[2.0ex]
& \; \left.  \left. \left.
+ z \, \left( \frac{1}{H_A - E - \hbar \omega_L} \right) \, z 
\right| \phi \right>
\right) \,,
\end{align}
which is a matrix element that involves only 
the atomic Green function, and is evaluated on the 
atomic reference state $| \phi \rangle$.
With the identification~\eqref{IL} of the laser intensity,
the second-order ac Stark shift
can be expressed in terms of the dipole polarizability 
$\alpha(\omega)$, which we write as
\begin{subequations}
\label{deltaE2}
\begin{align}
\delta E^{(2)} =& \; - \frac{I_L}{2 \epsilon_0 c} \, \alpha(\omega_L) \,,
\\[0.133ex]
\label{defpol}
\alpha(\omega) =& \;
\frac{e^2}{3} \, \sum_\pm 
\left< \phi \left| x^i \, 
\left( \frac{1}{H_A - E \pm \hbar \omega_L} \right) \, x^i 
\right| \phi \right> \,,
\end{align}
\end{subequations}
for a radially symmetric atomic reference state $|\phi \rangle$
which, in general, is the ground state.
The $x^i$ are the Cartesian components of the electron 
coordinate (the summation convention is used for $i$), 
while the sum over the two terms with both signs 
$\pm \omega$ will be encountered several times 
in the following; it is denoted by the symbol 
$\sum_\pm$ here.

The polarizability of an atom 
can be written as the sum of ``positive-frequency''
and ``negative-frequency'' 
components $\alpha_+(\omega_L)$ and $\alpha_-(\omega_L)$,
according to the formulas
\begin{subequations}
\label{osc_strength}
\begin{align}
\alpha(\omega_L) =& \; \alpha_+(\omega_L) + \alpha_-(\omega_L) \,,
\\[0.133ex]
\alpha_+(\omega_L) =& \; 
\frac{e^2}{3} \, 
\left< \phi \left| x^i \, 
\left( \frac{1}{H_A - E + \hbar \omega_L + \ii \, \epsilon} \right) \, x^i 
\right| \phi \right> \,,
\\[0.133ex]
\alpha_-(\omega_L) =& \;
\frac{e^2}{3} \,
\left< \phi \left| x^i \,
\left( \frac{1}{H_A - E - \hbar \omega_L - \ii \, \epsilon} \right) \, x^i  
\right| \phi \right> \,,
\end{align}
where the infinitesimal imaginary parts are introduced 
in accordance with the paradigm that the poles of the 
polarizability, as a function of $\omega_L$, have to be 
located in the lower half of the complex plane.
This is consistent with the fact that the polarizability corresponds
to a causal (retarded) Green function.
We take note of the identity
\begin{equation}
\label{dirac}
\frac{1}{x - \ii \epsilon} = \mbox{(P.V.)}
\frac{1}{x} + \ii \, \pi \, \delta(x) \,.
\end{equation}
We the help of this relation, one easily shows that 
the resonant, tree-level imaginary part of the polarizabilty 
${\rm Im}\left[ \alpha_R(\omega_L) \right]$
is an odd function of the driving frequency $\omega_L$,
\begin{align}
\label{ImAlphaR}
{\rm Im}\left[ \alpha_R(\omega_L) \right] =& \; 
\frac{\pi \, e^2}{3} \, 
\sum_{n,i} \left( \left| \left< \phi \left| x^i \right| \phi_n \right> \right|^2 \, 
\delta(E_n - E - \hbar \omega_L) 
\right.
\nonumber\\[0.133ex]
& \; - \left. 
\sum_{n,i} \left| \left< \phi \left| x^i \right| \phi_n \right> \right|^2 \, 
\delta(E_n - E + \hbar \omega_L) \right)
\nonumber\\[0.133ex]
=& \; 
\frac{\pi}{2} \, \sum_n \frac{f_{n0} \,
\delta(E_n - E - \hbar \omega_L)}{E_n - E} -
(\omega_L \leftrightarrow -\omega_L) \,,
\end{align}
\end{subequations}
where $E_n - E$ is the resonant frequency for the excitation of the ground state
to the excited virtual state $| \phi_n \rangle$ of energy $E_n$, 
and we have implicitly defined the dipole oscillator strength~$f_{n0}$
according to Ref.~\cite{YaBaDaDr1996}.
The second-order contribution to the polarizability 
${\rm Im}\left[ \alpha_R(\omega_L) \right]$
is in itself gauge invariant,
and it constitutes a sum of resonant peaks,
which are relevant as the laser frequency is tuned across 
an atomic resonance. Note that the contribution
${\rm Im}\left[ \alpha_R(\omega_L) \right]$
corresponds to the cutting of the following Feynman diagram,
\begin{center}
\includegraphics[width=0.3\linewidth]{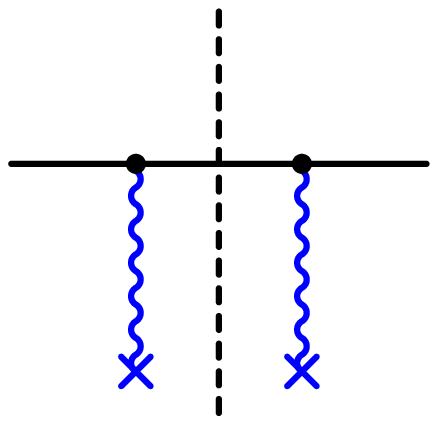}
\end{center}
according to the Cutkosky rules~\cite{ItZu1980}.
Wiggly lines denote photons (here, the interactions with the 
photons of the laser field).

\begin{figure}[t!]
\begin{center}
\begin{minipage}{1.0\linewidth}
\includegraphics[width=0.81\linewidth]{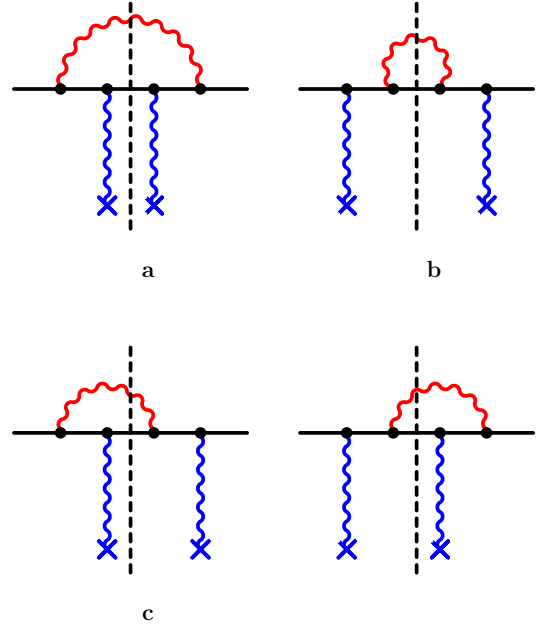}
\caption{\label{fig1} (Color online.) Feynman diagrams contributing to the
imaginary part of the polarizability in length gauge. 
Wiggly lines denote photons, straight lines denote the bound electron.
A laser photon
(interactions with the external laser field are denoted with a cross) is
absorbed from the laser field, while the self-energy loop describes the
self-interaction of the atomic electrons. When the virtual state denoted by the
internal line becomes resonant with the reference state, a pole is generated in
the integration over the degrees of freedom of the virtual photon (Cutkosky
rules~\cite{ItZu1980}, indicated by the vertical dashed lines).}
\end{minipage}
\end{center}
\end{figure}

However, there is an additional contribution
to the imaginary part of the polarizability,
created by the insertion of a virtual photon into the 
second-order diagram, as shown in Fig.~\ref{fig1}.
The terms with paired interactions with the 
laser and the radiation field, in the fourth order,
are given by
\begin{align}
& \delta E^{(4)} =
-\left< H_I \, G'(E_0)  \, H_L \, 
G'(E_0) \, H_L \, 
G'(E_0) \, H_I \right> 
\nonumber\\[0.133ex]
& \; -\left< H_L \, G'(E_0)  \, H_I \, 
G'(E_0) \, H_I \, G'(E_0) \, H_L \right> 
\nonumber\\[0.133ex]
& \; -2 \, \left< H_I \, G'(E_0) \, H_L \, 
G'(E_0) \, H_I \, G'(E_0) \, H_L \right> \,,
\end{align}
where $G'(E_0) = [1/(H_0 - E_0)']$ is the reduced Green function.
Written in this form, the terms correspond to the entries in Fig.~\ref{fig1},
(a)~and~(b), and the last term to the sum of (c) and~(d), while the photon loop
involves a photon in the mode $\vec k \, \lambda$.  The propagator denominators
in Fig.~\ref{fig1}(a) in the ``outer'' legs of the electron line read as
$H-E+\omega_{\vec k \, \lambda}$ 
(twice) because the spontaneously emitted photon is part of the
virtual state of atom$+$field.  By contrast, the propagator denominators in the
``outer'' lines in Fig.~\ref{fig1}(b) read as $H-E-\omega_L$ (twice) because the
spontaneously emitted photon is not present. In all cases, one photon is
``taken from'' the laser field.  For Figs.~\ref{fig1}(c) and~(d), we have a
mixed configuration with two propagators, with denominators $(H-E+\omega)$ and
$(H-E-\omega)$.

In the spirit of Ref.~\cite{CrTh1984},
the diagram~(a) in Fig.~\ref{fig1} entails the 
following energy shift,
\begin{align}
\delta E_a =& \; -e^4 
\sum_{\vec k  \lambda}
\frac{\hbar \omega_L}{2 \epsilon_0 \calV_L}
\frac{\hbar \omega_{\vec k \, \lambda}}{2 \epsilon_0 \calV}
 \left< \phi_0 \left| 
(\hat\epsilon_{\vec k\, \lambda} \cdot \vec r) \,
\left( a_{\vec k \, \lambda}^+ + a_{\vec k \, \lambda} \right) \
\right. \right. 
\nonumber\\[0.133ex]
& \; \times \frac{1}{H_0 - E_0} \,
z \, (a_L^+ + a_L) \, 
\frac{1}{H_0 - E_0} \, 
z \, (a_L^+ + a_L) \, 
\nonumber\\[0.133ex]
& \; \left.  \left. \times 
\frac{1}{H_0 - E_0} \, 
(\hat\epsilon_{\vec k\, \lambda} \cdot \vec r) \,
\left( a_{\vec k \, \lambda}^+ + a_{\vec k \, \lambda} \right) \
\right| \phi_0 \right> \,.
\end{align}
We now use the matching condition~\eqref{match} 
and calculate the sum over polarizations
using Eq.~\eqref{sumpol}.
Isolating the terms which describe the absorption 
of a photon from the laser field
and emission into the field mode $\vec k \, \lambda$,
we obtain the following expression,
\begin{align}
\delta E_a \sim& \; -e^4 
\int \frac{\dd^3 k}{(2\pi)^3} 
\, \frac{n_L \hbar \omega_L}{2 \epsilon_0 \calV_L}
\, \frac{\hbar \omega_{\vec k \, \lambda}}{2 \epsilon_0}
\left( \delta^{ij} - \frac{k^i\,k^j}{\vec k^{\,2}} \right) 
\nonumber\\[0.133ex]
& \; \times 
\left< \phi \left| z \frac{1}{H_A - E + \hbar \omega_{\vec k \lambda}} 
x^i \frac{1}{H_A - E - \hbar \omega_L + \hbar \omega_{\vec k \lambda}} 
\right.
\right.
\nonumber\\[0.133ex]
& \; 
\left. 
\left. 
\times 
x^j \, \frac{1}{H_A - E + 
\hbar \omega_{\vec k \, \lambda}} \, z \, \right| \phi \right> \,.
\end{align}
The $\sim$ sign implies that the terms which involve the 
virtual state $| \phi, n_L - 1, 1_{\vec k \lambda}\rangle$
still need to be isolated.
We have traced out the photon degrees of freedom.
We can isolate the imaginary contribution 
created by the virtual state $| \phi, n_L - 1, 1_{\vec k \lambda}\rangle$,
with $\omega_{\vec k \lambda} = \omega_L$, as follows,
\begin{align}
& \; \frac{1}{H_A - E - \hbar \omega_L + \hbar \omega_{\vec k \, \lambda} } 
\nonumber\\[0.133ex]
& \to 
\frac{1}{H_A - E - \hbar \omega_L + \hbar \omega_{\vec k \, \lambda} - \ii \epsilon} 
\nonumber\\[0.133ex]
& \; \to
\frac{ | \phi \rangle \, \langle \phi | }%
{- \hbar \omega_L + \hbar \omega_{\vec k \, \lambda} - \ii \epsilon} 
\to 
\frac{\ii \, \pi}{\hbar} \, \delta( \omega_{\vec k \, \lambda} - \omega_L ) \;
| \phi \rangle \, \langle \phi | \,.
\end{align}
Here, an infinitesimal imaginary part has been introduced 
according to the Feynman prescription for the positive-energy,
virtual atomic states.
Keeping track of the prefactors, one finally obtains
\begin{align}
\ii \, {\rm Im}(\delta E_a)
=& \; - \ii \frac{I_L}{2 \epsilon_0 c} \, \Biggl\{  
\frac{\omega^3_L }{6 \pi \epsilon_0 c^3} 
\\[0.133ex]
& \; 
\times \left( \frac{e^2}{3} \left< \phi \left| x^i \, 
\frac{1}{H_A - E + \hbar \omega_L} \, x^i
\right| \phi \right> \right)^2 \Biggr\} \,,
\nonumber
\end{align}
where the Einstein summation convention is used for 
repeated indices. 
The term in curly brackets, 
upon comparison with Eq.~\eqref{deltaE2},
can be identified as the imaginary 
part of the polarizability 
associated with diagram~(a) of Fig.~\ref{fig1}.
The imaginary part ${\rm Im}(\delta E_a)$ of the 
energy shift associated with Fig.~\ref{fig1}(a)
is negative, as it should be for a decay process,
while the imaginary part of the 
polarizability is positive [see the overall minus sign 
in Eq.~\eqref{deltaE2}].

Diagram~(b) in Fig.~\ref{fig1} 
generates an imaginary part in much the same way,
but the ``outer'' virtual states have one 
virtual laser photon less than the 
intermediate state which generates the imaginary part,
hence
\begin{align}
\ii \, {\rm Im}(\delta E_b)
=& \; - \ii \frac{I_L}{2 \epsilon_0 c} \, \Biggl\{ 
\frac{\omega^3_L }{6 \pi \epsilon_0 c^3} 
\\[0.133ex]
& \; 
\times \left( 
\frac{e^2}{3} 
\left< \phi \left| x^i \, 
\frac{1}{H_A - E - \hbar \omega_L} \, x^i
\right| \phi \right> \right)^2 \Biggr\} \,.
\nonumber
\end{align}
Finally, the sum of diagrams~(c) and~(d) 
generates the following expression,
\begin{align}
\ii \, {\rm Im}(\delta E_{c+d})
=& \; - \ii \frac{I_L}{2 \epsilon_0 c} \, 
\left\{ \frac{e^4 \, \omega^3_L }{27 \pi \epsilon_0 c^3} 
\right.
\nonumber\\[0.133ex]
& \; \times 
\left< \phi \left| x^i \, 
\frac{1}{H_A - E + \hbar \omega_L} \, x^i \right| \phi \right> \,
\nonumber\\[0.133ex]
& \; \times \left. 
\left< \phi \left| x^j \, 
\frac{1}{H - E - \hbar \omega_L} \, x^j \right| \phi \right>  
\right\} \,.
\end{align}
The end result for the imaginary part of the 
polarizability, generated by the fourth-order diagrams
in Fig.~\ref{fig1}, reads as
\begin{align}
\label{endres}
\ii \, {\rm Im}(\delta E^{(4)})
=& \; \ii \, {\rm Im}(\delta E_{a+b+c+d})
= - \ii \frac{I_L}{2 \epsilon_0 c } \, 
\left\{  \frac{\omega^3_L}{6 \pi \epsilon_0 c^3} \right.
\nonumber\\[0.133ex]
& \; \times 
\left( 
\frac{e^2}{3} 
\left< \phi \left| x^i \, 
\frac{1}{H_A - E + \hbar \omega_L} \, x^i \right| \phi \right> 
\right.
\nonumber\\[0.133ex]
& \; \left. \left. +
\frac{e^2}{3} 
\left< \phi \left| x^j \, 
\frac{1}{H_A - E - \hbar \omega_L} \, x^j \right| \phi \right>  
\right)^2 \right\} \,.
\end{align}
After matching with Eq.~\eqref{deltaE2},
it can be summarized in the following, compact result,
\begin{subequations}
\label{mainres}
\begin{equation}
\label{mainresa}
{\rm Im}[ \alpha(\omega_L) ] =
{\rm Im}[ \alpha_R(\omega_L) ] +
\frac{\omega^3_L}{6 \pi \epsilon_0 c^3} \, 
[ \alpha(\omega_L) ]^2 \,,
\end{equation}
where ${\rm Im}[ \alpha_R(\omega_L) ]$ is the sum of the
resonance peaks, according to Eq.~\eqref{ImAlphaR},
and in the last term, the square of the 
polarizability is calculated using 
propagators without any added ``damping terms''
for the virtual state energies, i.e.,
under the identification 
$[ \alpha(\omega_L) ]^2 \equiv {\rm Re}[ \alpha(\omega_L) ]^2$
[see Eq.~\eqref{endres}].
The result~\eqref{mainres} is manifestly odd in the 
argument~$\omega_L$, as it should be [see Eq.~\eqref{one}].

Let us dwell on the precise form of this result a little more.
As shown below in Eq.~\eqref{alpha3},
the second term on the right-hand side of Eq.~\eqref{mainresa}
is a correction of relative order $\alpha_{\rm QED}^3$,
where $\alpha_{\rm QED}$ is the fine-structure constant.
Furthermore, the contribution given in Eq.~\eqref{ImAlphaR}
can be identified, in terms of a quantum electrodynamic 
formalism, as the ``tree-level'' contribution to the 
imaginary part, or, as the imaginary part of the 
``tree-level'' polarizability $\alpha_{\rm TL}(\omega)$ itself.
If we absorb in our definition of $\alpha_{\rm TL}(\omega)$
relativistic corrections to the 
dipole polarizability of relative order $\alpha_{\rm QED}^2$
(see Ref.~\cite{Ya2003})
and all radiative corrections without cuts in the 
self-energy photon lines
(of relative order $\alpha_{\rm QED}^3$, see 
Refs.~\cite{BeSa1957,ItZu1980,BeJa1986,HaEtAl2006}), 
then we can write Eq.~\eqref{mainresa} alternatively as 
\begin{equation}
\label{mainresb}
\frac{{\rm Im}[ \alpha(\omega) ]}{\alpha(\omega)} = 
\frac{{\rm Im}[ \alpha_{\rm TL}(\omega) ]}{\alpha_{\rm TL}(\omega)} +
\frac{\omega^3}{6 \pi \epsilon_0 c^3} \,
\alpha_{\rm TL}(\omega)
+\calO(\alpha_{\rm QED}^4) \,.
\end{equation}
We can ``sum'' the second term into a denominator,
\begin{equation}
\label{mainresc}
{\rm Im}[ \alpha(\omega) ] \approx
{\rm Im}\left[ 
\frac{\alpha_{\rm TL}(\omega)}{
1 - \ii \, \dfrac{\omega^3}{6 \pi \epsilon_0 c^3} \, 
\alpha_{\rm TL}(\omega) } \right] \,.
\end{equation}
\end{subequations}
The latter functional form is in agreement with various results
for polarizabilities of nanoparticles (not atoms)
and other structures, corrected for ``radiation
damping'' [see Eq.~(5) of Ref.~\cite{CaEtAl2006optcom},
Eq.~(27) of Ref.~\cite{AlEtAl2010prb},
Eq.~(5) of Ref.~\cite{MaGA2010},
Eq.~(4) of Ref.~\cite{MaGA2012},
and Eq.~(34) of Ref.~\cite{MeEtAl2013}].
The term from the denominator in Eq.~\eqref{mainresc}
has the required functional from for ``radiation damping''
due to the Abraham--Lorentz radiation reaction force
[see Eq.~(1) and~(10) of Ref.~\cite{LaDKJe2012cejp},
and Refs.~\cite{Bo1969friction,ZSGrNo2004}].
The tree-level term is identified with the
``damping by absorption'' of the nanoparticle,
\begin{equation}
{\rm Im}[ \alpha_{\rm TL}(\omega) ] \approx
{\rm Im}[ \alpha(\omega) ] -
\dfrac{\omega^3}{6 \pi \epsilon_0 c^3} \,
[ \alpha_{\rm TL}(\omega) ]^2 \,.
\end{equation}
or, as the 
``non-radiative'' imaginary part in the
sense of Ref.~\cite{CaEtAl2006optcom}.
Note that the ``tree-level'' for a nanoparticle term is denoted 
as $\alpha_0(\omega)$ in Ref.~\cite{CaEtAl2006optcom},
which however, for an atom, would be associated
with the ``monopole'' polarizability which is vanishing
(canonically, one denotes by $\alpha_\ell(\omega)$ the 
$2^\ell$-pole polarizability of an atom,
see Ref.~\cite{LaDKJe2010pra}).

A possible interconnection
of the imaginary part of the polarizability with the
square of the polarizability has previously been formulated in
Refs.~\cite{ZSGrNo2004} and~\cite{NoHe2012}.  A term with the square of the
polarizability is added to the imaginary part of the polarizability in
Eqs.~(G2) and~(G3) of Ref.~\cite{ZSGrNo2004} [see also Eq.~(49) of
Ref.~\cite{Bo1969friction}.  Within QED, the term with the square of the
polarizability is identified as being due to a one-loop correction, while the
resonant term is added to the results given in
Refs.~\cite{Bo1969friction,ZSGrNo2004,NoHe2012} and constitutes a tree-level
contribution.

%
%
\section{Velocity Gauge}
\label{sec3}

The question of the choice of gauge in electrodynamic interactions has been
raised over a number of decades~\cite{PoZi1959}, 
with a particularly interesting analysis being
presented in Ref.~\cite{ScBeBeSc1984}.  In general, and in accordance with the
now famous remark by Lamb on p.~268 of Ref.~\cite{La1952}, the wave
function of a bound state therefore preserves its
probabilistic interpretation only in the length gauge, and it is this gauge
which should be used for off-resonance excitation
processes~\cite{Ko1978prl,ScBeBeSc1984}. 
According to Table~1 of of Ref.~\cite{PoTh1978},
the dipole matrix elements for dipole transitions 
from state $|i\rangle$ to state $|f\rangle$ in the 
dipole approximation are related by the 
\begin{equation}
\left< f \left| \frac{e}{m} \vec p \cdot \vec A \right| i \right> 
= \ii \, \frac{E_f - E_i}{\hbar \omega} 
\left< f \left| e \vec r \cdot \vec E \right| i \right>  \,,
\end{equation}
as can be shown with the help of the commutation 
relation $\vec p = \frac{\ii m}{\hbar} [H_A, \vec r]$.
The two forms are equivalent only at resonance 
$\hbar \omega = E_f - E_i$.
In quantum dynamic calculations, which often involve 
nonresonant driving of atomic transitions,
according to Lamb's remark on p.~268 of Ref.~\cite{La1952}, 
``a closer examination shows that the usual interpretation
of probability amplitudes is valid only in the former 
[the length] gauge, and no additional 
factor [$(E_f - E_i)/(\hbar \omega)$] occurs.''
Namely, the momentum operator $\vec p = -\ii \hbar \vec \nabla$
describes the mechanical momentum of the 
matter wave in the length gauge, where the 
kinetic and canonical momentum operators assume the 
same form~\cite{ScBeBeSc1984,JeKe2004aop}. In the velocity gauge, the canonical
momentum $\vec p - e\, \vec A$ assumes the role of the 
conjugate momentum of the position operator in the equations of motion.
One can also argue that the electric field
used in the length-gauge interaction is gauge invariant,
while the vector potential in the velocity-gauge 
term is not~\cite{Ko1978,ScBeBeSc1984,JeEvHaKe2003,JeKe2004aop,EvJeKe2004,JeEvKe2005}.
However, for processes such as energy
shifts, where the initial and final states have the same energy and the sum of
the exchanged photon energies need to add up to zero, both gauges should lead
to the same results (see Sec.~3.3 of Ref.~\cite{JeKe2004aop},
and Refs.~\cite{Je2004rad,MeJhHi1990,Ha1992,Je2015rapid}).
While limits of gauge invariance have recently been discussed 
in Ref.~\cite{Re2013}, it is a nontrivial exercise to check 
the gauge invariance of the imaginary part of the polarizability.
Similar questions have been discussed in the context of the 
QED radiative correction to laser-dressed states (Mollow spectrum,
see Refs.~\cite{JeEvHaKe2003,JeKe2004aop,EvJeKe2004,JeEvKe2005}).

In the velocity gauge, 
the interactions are formulated in terms 
of the vector potentials $\vec A_L$ and 
$\vec A_I$, which denote the laser field 
and the interaction with the 
quantized radiation field.
We denote the corresponding interaction 
Hamiltonians, in the dipole approximation,
by the lowercase letters $h_L$ and $h_I$ and 
write
\begin{subequations}
\begin{align}
\vec A_L =& \; \hat e_z \, \sqrt{\frac{\hbar}{2 \omega_L \epsilon_0 \calV_L}} \,
\left( \ii a_L - \ii a_L^+ \right) 
= \hat e_z \, A_L \,,
\\[0.133ex]
\vec A_I =& \; \sum_{\vec k \, \lambda}
\hat\epsilon_{\vec k\, \lambda} \,
\sqrt{\frac{\hbar}{2 \omega_{\vec k \, \lambda} \epsilon_0 \calV}} \,
\left( \ii a_{\vec k \, \lambda} - \ii a^+_{\vec k \, \lambda} \right) 
\\[0.133ex]
h_L =& \; -\frac{e \, A_L \, p_z}{m}  \,,
\qquad
h_I = -\frac{e \, \vec A_I \cdot \vec p}{m}  \,,
\end{align}
\end{subequations}
From the seagull term, we isolate the 
terms which involve at least one interaction 
with the laser mode and write
\begin{align}
h_S =& \; e^2 \, \frac{(\vec A_L + \vec A_I)^2}{2 m} 
\to h_{LI} + h_{LL} \,,
\nonumber\\[0.133ex]
\qquad h_{LI} =& \; e^2 \, \frac{\vec A_L \cdot \vec A_I}{m} \,,
\qquad
\qquad h_{LL} = e^2 \, \frac{\vec A_L^2}{2 m} \,.
\end{align}
In the length gauge,
the second-order shift in the laser field
is given by Eq.~\eqref{deltaE2}.
In the velocity gauge, 
one obtains an additional term, as follows,
\begin{equation}
\delta E^{(2)} = \langle h_{LL} \rangle
-\left< h_L \, \frac{1}{H_0 - E_0}  \, h_L \right> \,.
\end{equation}
We trace out the laser photon degrees of freedom and obtain
\begin{align}
\label{deltaE2V}
\delta E^{(2)} =& \;
-e^2 \, \frac{n_L \hbar \omega_L}{2 \epsilon_0 \calV_L}
\left( \frac{1}{\omega_L^2} \,
\left< \phi \left| \frac{p_z}{m} \,
\left( \frac{1}{H_A - E + \hbar \omega_L} \right) \, \frac{p_z}{m}
\right.  \right.  \right.
\nonumber\\[2.0ex]
& \; \left.  \left. \left.
+ \frac{p_z}{m} \, \left( \frac{1}{H_A - E - \hbar \omega_L} \right) \, 
\frac{p_z}{m} \right| \phi \right>
- \frac{1}{m \, \omega_L^2} \right) \,,
\end{align}
where $p_z$ is the $z$ component of the momentum 
operator. One might conclude that this form is 
manifestly different from Eq.~\eqref{deltaE2}.
Using the operator identity 
\begin{equation}
\frac{p^i}{m} = \frac{\ii}{\hbar} \, [H - E + \hbar \omega_L, r^i ] \,,
\end{equation}
one may show the relation
\begin{align}
\label{vellen}
& \frac13 \, \left< \phi \left| \frac{p^i}{m} \,
\frac{1}{H - E + \hbar\omega_L} \, \frac{p^i}{m} \right| \phi \right> =
\frac{1}{2m}
\\[0.133ex]
& \; - \frac{\omega_L}{3 \hbar} \,
\left< \phi \left| \vec r^{\,2} \right| \phi \right> + 
\frac{\omega_L^2}{3} \, \left< \phi \left| 
x^i \, \frac{1}{H - E + \hbar \omega_L} \, x^i \right| \phi \right> \,,
\nonumber
\end{align}
where $\vec r^{\,2} = x^i \, x^i$.
With the help of this relation, and assuming a
spherically symmetric atomic reference state $|\phi\rangle$, 
one may confirm that the second-order shifts~\eqref{deltaE2V}
and~\eqref{deltaE2} are equal.

\begin{figure}[t!]
\begin{center}
\begin{minipage}{1.0\linewidth}
\includegraphics[width=0.71\linewidth]{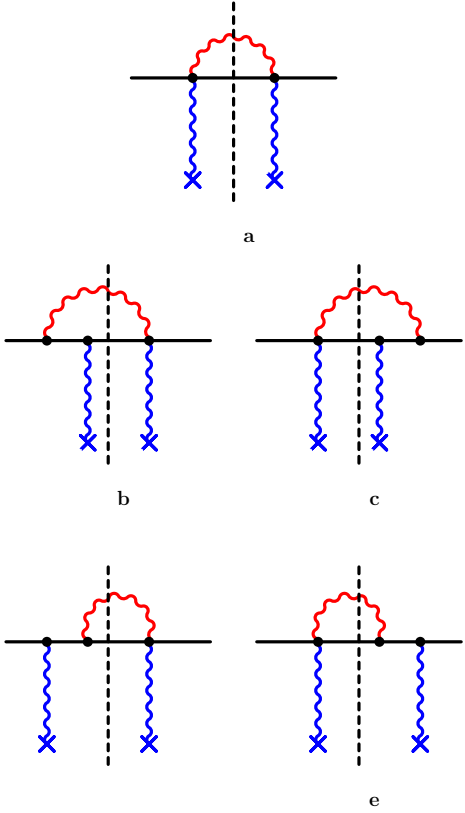}
\caption{\label{fig2} (Color online.) 
Additional Feynman diagrams involving the 
seagull graph which have to be considered in the 
velocity gauge, in order to 
show the gauge invariance of the 
imaginary part of the polarizability.}
\end{minipage}
\end{center}
\end{figure}

In the additional diagrams which persist in the velocity gauge (see
Fig.~\ref{fig2}), the seagull graph enters ``in disguise'', with the virtual
photon of the self-energy loop and the laser photon emerging from the same
vertex.  We may anticipate that diagram~(a) in Fig.~\ref{fig2} generates a
constant term, while diagrams~(b) and~(c) generate terms with one propagator,
with a propagator denominator of the form $H-E+\omega_L$ (the spontaneously
emitted photon is present in the ``outer'' lines).  By contrast, the diagrams
in Fig.~\ref{fig2}~(d) and~(e) generate propagator denominators $H-E-\omega_L$
for the internal line of the diagram.

From the diagram in Fig.~\ref{fig2}(a), we have
the energy shift
\begin{equation}
\delta \calE_a =
-\left< \phi_0 \left| h_{LI} \, \frac{1}{H_0 - E_0}  \, h_{LI} 
\right| \phi_0 \right> \,.
\end{equation}
After tracing out the photon degrees of freedom,
the imaginary part can be extracted as follows,
\begin{equation}
{\rm Im}( \delta \calE_a ) = - \ii \, \frac{I_L}{2 \epsilon_0 c} \, 
\frac{e^4}{6 \pi \epsilon_0 \, c^3 \, m^2 \omega_L} \,.
\end{equation}
From the diagram in Fig.~\ref{fig2}(b) and (c), we have
the energy shift
\begin{align}
\delta \calE_{b+c} =& \;
2 \, \left< \phi_0 \left| h_I \,
\frac{1}{H_0 - E_0} \, h_L
\frac{1}{H_0 - E_0} \, h_{LI}
\right| \phi_0 \right> \,.
\end{align}
The imaginary part of the energy, generated by 
the diagrams Fig.~\ref{fig2}(b) and (c), can conveniently 
be written as follows,
\begin{align}
{\rm Im}( \delta \calE_{b+c} ) =& \;
\ii \, \frac{I_L}{2 \epsilon_0 c} \, 
\frac{e^2}{3 \pi \epsilon_0 c^3 m} \, \omega_L \,
\\[0.133ex]
& \; \times \left( \frac{e^2}{3 \, \omega_L^2} \,
\left< \phi \left| \frac{p^i}{m} \, \frac{1}{H - E + \hbar \omega_L} \,
\frac{p^i}{m} \right| \phi \right> \right) \,.
\nonumber
\end{align}
Finally, from the diagrams in Fig.~\ref{fig2}(d) and (e), we have
the energy shift
\begin{align}
\delta \calE_{d+e} =& \;
2 \, \left< \phi_0 \left| h_L \,
\frac{1}{H_0 - E_0} \, h_I
\frac{1}{H_0 - E_0} \, h_{LI}
\right| \phi_0 \right>
\end{align}
where the sequence of the dipole coupling 
$h_L$ and $h_I$ is reversed in comparison to 
the diagrams in Fig.~\ref{fig2}(b) and~(c).
The imaginary part of the energy shift is the same
up to a sign change in the frequency term in the
propagator denominator,
\begin{align}
{\rm Im}( \delta \calE_{d+e} ) =& \;
\ii \, \frac{I_L}{2 \epsilon_0 c} \,
\frac{e^2}{3 \pi \epsilon_0 c^3 m} \, \omega_L \,
\\[0.133ex]
& \; \times \left( \frac{e^2}{3 \, \omega_L^2} \,
\left< \phi \left| \frac{p^i}{m} \, \frac{1}{H - E - \hbar \omega_L} \,
\frac{p^i}{m} \right| \phi \right> \right) \,.
\nonumber
\end{align}
The diagrams in Fig.~\ref{fig2} all involve
at least one occurrence of the 
combined laser-quantized-field 
seagull term $h_{LI}$.

Finally, we can write the additional imaginary part
due to the seagull terms, denoted by ${\rm Im}(\delta \calE_2) = 
{\rm Im}( \delta \calE_{a+b+c+d+e} )$, 
in the velocity gauge as follows,
\begin{align}
{\rm Im}(\delta \calE_2) =& \; 
- \ii \, \frac{I_L}{2 \epsilon_0 c} \, \frac{1}{6 \pi \epsilon_0 c^3} \,
\frac{e^4}{m^2 \, \omega_L} 
\nonumber\\[0.133ex]
& \; + \ii \, 
\frac{I_L}{2 \epsilon_0 c} \, 
\frac{e^2}{3 \pi \epsilon_0 c^3 m} \, \omega_L \,
\nonumber\\[0.133ex]
& \; \times \left[ \frac{e^2}{3 \, \omega_L^2} \, \left(
\left< \phi \left| \frac{p^i}{m} \, \frac{1}{H - E + \hbar \omega_L} \,
\frac{p^i}{m} \right| \phi \right>  
\right.  \right.
\nonumber\\[0.133ex]
& \; \left. \left. +
\left< \phi \left| \frac{p^i}{m} \, \frac{1}{H - E - \hbar \omega_L} \,
\frac{p^i}{m} \right| \phi \right> \right) \right]
\nonumber\\[0.133ex]
=& \; \ii \, \frac{I_L}{2 \epsilon_0 c} \, 
\frac{e^2}{6 \pi \epsilon_0 c^3 m} \,
\left[ \frac{e^2}{m \, \omega_L} + 2 \, \omega_L \, 
\alpha(\omega_L) \right] \,.
\end{align}
We have used the identity~\eqref{vellen}.
To this result we have to add the contribution 
of the diagrams in Fig.~\ref{fig1}, this time 
evaluated with the dipole Hamiltonians 
replaced by their velocity-gauge counterparts
($H_L \to h_L$ and $H_I \to h_I$),
\begin{align}
\ii \, {\rm Im}( \delta \calE_1 ) =& \; - \ii
\frac{I_L}{2 \epsilon_0 c} \, \frac{1}{6 \pi \epsilon_0 c^3} \, \frac{1}{\omega_L} \,
\nonumber\\[0.133ex]
& \; \times \left[ \frac{e^2}{3} \,
\sum_\pm \left< \phi \left| \frac{p^i}{m} \,
\frac{1}{H - E \pm \hbar \omega_L} \, \frac{p^i}{m} \right| \phi \right> 
\right]^2
\end{align}
Again, with the help of~\eqref{vellen},
we can write this as
\begin{align}
\ii \, {\rm Im}( \delta \calE_1 ) =& \; - \ii
\frac{I_L}{2 \epsilon_0 c} \, 
\frac{e^2}{6 \pi \epsilon_0 c^3 m} \,
\left\{ \frac{e^2}{m \, \omega_L} +
2 \, \omega_L \, \alpha(\omega) \right\}
\nonumber\\[0.133ex]
& \; - \ii \, \frac{I_L}{2 \epsilon_0 c} \, 
\frac{1}{6 \pi \epsilon_0 c^3} \,
\omega_L^3 \, [\alpha(\omega_L)]^2 \,.
\end{align}
Finally, one obtains in the velocity gauge
\begin{equation}
\label{vel_gauge}
{\rm Im}(\delta \calE_1) +
{\rm Im}(\delta \calE_2) =
- \frac{I_L}{2 \epsilon_0 c} \, 
\frac{1}{6 \pi \epsilon_0 c^3} \,
\omega_L^3 \, [\alpha(\omega_L)]^2 \,,
\end{equation}
confirming the result~\eqref{endres} once more
[the oscillator strength in Eq.~\eqref{osc_strength}
is in itself gauge invariant, and the 
resonant contribution~\eqref{ImAlphaR} needs to be 
added to the off-resonant contribution~\eqref{vel_gauge}].

%
%
\section{Conclusions}
\label{sec4}

Our main result concerns the imaginary part of the 
atomic polarizability, which is the sum of a 
resonant term ${\rm Im}[ \alpha_R(\omega_L) ]$ and of 
an off-resonant driving of an 
atom with the dynamic dipole polarizability $\alpha(\omega)$.
According to Eq.~\eqref{mainres}, the result is given 
as follows,
\begin{equation}
\label{repeat}
{\rm Im}[ \alpha(\omega_L) ] = 
{\rm Im}[ \alpha_R(\omega_L) ] +
\frac{\omega^3_L}{6 \pi \epsilon_0 c^3} \, 
{\rm Re}[ \alpha(\omega_L) ]^2 \,.
\end{equation}
The clarification of the functional dependence of the 
nonresonant contribution to the 
imaginary part for small $\omega$ has been a matter 
of discussion in the past 
(see Chap.~XXI of Ref.~\cite{Me1962vol2} and 
Refs.~\cite{LaDKJe2012prl,JeLaDKPa2015}),
with both a linear dependence on $\omega$ and 
an $\omega^3$ dependence being discussed as candidate formulas.
Initially, one would assume that the 
$\omega^3$ dependence is favored in the ``length gauge'',
whereas the linear term is obtained in the 
``velocity gauge'' (see the discussion in Ref.~\cite{LaDKJe2012prl}),
but the formulation presented here in terms of 
the imaginary part of a (necessarily gauge-invariant) 
fourth-order energy shift removes the ambiguity.
So, our result~\eqref{repeat} settles the question.
While the calculations have been described for a single-electron
atom (hydrogen atom), the results hold more generally
because one may simply add, in the calculation of the
polarizability, the dipole couplings of the other electrons,
according to the replacement $x^i \to \sum_a x^i_a$
where $a$ denotes the subscript of the electron.
The full inclusion of all diagrams
given in Figs.~\ref{fig1} and~\ref{fig2}
is crucial in confirming the gauge invariance
(see Secs.~\ref{sec2} and~\ref{sec3}).

Our results indicate that the imaginary part of the 
atomic polarizability 
cannot be obtained on the basis of a replacement 
in the propagator denominators of the 
polarizability function,
\begin{align}
\label{replacement}
\alpha(\omega) =& \;
\frac{e^2}{3} \, \sum_\pm 
\frac{\left| \left< \phi \left| x^i \right| \phi_m \right> \right|^2}%
{E_m - E \pm \hbar \omega} 
\nonumber\\[0.133ex]
\to & \;
\frac{e^2}{3} \, \sum_\pm 
\frac{\left| \left< \phi \left| x^i \right| \phi_m \right> \right|^2}%
{E_m - \frac{\ii}{2} \, \Gamma_m - E \pm \hbar \omega}  \,.
\end{align}
Beyond tree level, the latter prescription
cannot possibly lead to a consistent result for the 
imaginary part of the polarizability,
no matter how one extrapolates the 
decay width $\Gamma_m \equiv \Gamma_m(\omega)$ off resonance.
However, we may at least observe that 
if the $\Gamma_m$ represent the one-photon 
decay widths of the virtual states, then the 
replacement~\eqref{replacement} generates 
the right order-of-magnitude 
for the imaginary part off resonance,
in agreement with Eq.~\eqref{repeat}. 

The specification of the real part of the 
square of the polarizability in Eq.~\eqref{repeat} 
serves to maintain the relevant symmetry:
Namely, ${\rm Im}[ \alpha(\omega_L) ]$ needs to be 
an odd function under a sign change of its argument.
However, the off-resonant (one-loop) contribution to the 
imaginary part of the polarizability is in itself 
a radiative correction and should thus be 
suppressed by powers of the fine-structure 
constant $\alphaQED$. Hence, to leading order in $\alphaQED$, 
we can leave out the additional specification of the 
real part in the second term in Eq.~\eqref{repeat}.
To put this statement into context, 
we switch to natural units with 
$\hbar = c = \epsilon_0 = 1$ and 
refer to Ref.~\cite{Be1947} where it is shown 
that the imaginary part
of the one-loop self-energy describes an 
imaginary energy contribution of order $\alphaQED^5 \, m$, 
where $\alphaQED$ is the fine-structure constant
[in contrast to $\alpha(\omega)$ which denotes the 
polarizability].
The radiative correction to the energy is of relative order 
$\alphaQED^3$ in comparison to the 
Schr\"{o}dinger energy; the latter is of order 
$\alphaQED^2 \, m$ in natural units.

Let us apply this program to the polarizability.
Taking into account that the Bohr radius is 
of order $1/(\alphaQED \, m)$ in natural units 
(i.e., about 137 times larger than the reduced Compton 
wavelength of the electron), 
we obtain the following order-of-magnitude estimates:
\begin{equation}
\label{alpha3}
\alpha(\omega) \sim \frac{\alphaQED}{(\alphaQED m)^2} \, 
\frac{1}{\alphaQED^2 m} = \frac{1}{(\alphaQED m)^3} \,.
\end{equation}
In order to perform the estimate for 
${\rm Im}[\alpha(\omega)] \sim m^{-3}$,
we need to take into account that a typical 
atomic driving frequency is of order $\omega_L \sim \alphaQED^2 \, m$,
and the ratio again is of the order of
\begin{equation}
\frac{{\rm Im}[\alpha(\omega)]}{\alpha(\omega)} \sim \alphaQED^3 \,,
\end{equation}
as it should be for a one-loop radiative correction
[here, ${\rm Im}[\alpha(\omega)]$ is restricted to the second term 
in Eq.~\eqref{repeat}].

Because of the general nature of the obtained 
result, it is useful to discuss the 
conversion to other unit systems.
In natural units (n.u.), with $\hbar = c = \epsilon_0 = 1$,
the result~\eqref{repeat} simplifies to
\begin{equation}
\label{mainres2}
\left. {\rm Im}[ \alpha(\omega) ] \right|_{\rm n.u.} =
\left. {\rm Im}[ \alpha_R(\omega) ] \right|_{\rm n.u.} +
\frac{\left. \omega^3 \right|_{\rm n.u.} }{6 \pi} \,
\left. [ \alpha(\omega) ]^2 \right|_{\rm n.u.} \,.
\end{equation}
Using the relation 
$e^2 = 4\pi\alphaQED$, we can write the 
dipole polarizability, in natural units, as follows,
\begin{align}
& \left.  \alpha(\omega) \right|_{\rm n.u.} 
= \frac{4 \pi \alphaQED}{\alphaQED^4} \, 
\nonumber\\[0.133ex]
& \;
\times 
\left\{ \frac{1}{3} \, \sum_\pm \left< \phi \left| \alphaQED \, m \, x^j \,
\frac{\alphaQED^2 m}{H - E \pm \omega_L} \, 
\alphaQED \, m \, x^j \right| \phi \right> \right\} \,.
\end{align}
Here, the sum over the signs $\pm$ denotes the 
two terms generated by virtual absorption and emission
of the laser photon, as in Eq.~\eqref{defpol}.
For an atom, all quantities in the term in curly 
brackets are of order unity; e.g., 
$\alphaQED \, m \, x^j = x^j/a_0 = \rho^j$ where
$a_0$ is the Bohr radius.
The numerical values (reduced quantities) 
associated with the polarizability, in 
atomic units (a.u.) and natural units (n.u.),
are thus related by
\begin{equation}
\left.  \alpha(\omega) \right|_{\rm n.u.} 
= \frac{4 \pi}{\alphaQED^3} \,
\left. \alpha(\omega) \right|_{\rm a.u.} \,.
\end{equation}
Atomic transition frequencies are measured 
in Hartrees in atomic units, and thus we have
\begin{equation}
\left. \omega_L \right|_{\rm n.u.} = 
\alphaQED^2 \, m \; \left. \omega_L \right|_{\rm a.u.} \,.
\end{equation}
So, in atomic units, our main result
given in Eq.~\eqref{mainres} and~\eqref{mainres2}
reads as 
\begin{equation}
\label{ImAlphaAU}
\left. {\rm Im}[\alpha(\omega)] \right|_{\rm a.u.} =
\left. {\rm Im}[\alpha_R(\omega)] \right|_{\rm a.u.} +
\frac{2 \alphaQED^3}{3} \, 
[ \left.  \omega^3 \; \alpha^2(\omega) ] \right|_{\rm a.u.} \,.
\end{equation}
The imaginary part of the 
polarizability enters the description of 
a number of dissipative processes, such as the 
quantum friction due to interaction with a bath
of black-body photons~\eqref{force1},
or, with a dielectric surface [see Eq.~\eqref{force2}].
The imaginary part of the polarizability 
is manifestly nonvanishing off resonance and, 
for small driving frequency $\omega$, is proportional
to $\omega^3$.

\section*{ACKNOWLEDGMENTS}

Helpful conversations with M.~De Kieviet are gratefully acknowledged.
The authors (U.D.J.) wish to acknowledge support from the National
Science Foundation (Grants PHY--1068547 and PHY--1403973)
and (K.P.) from the Polish National Science Center (NCN, Grant 2012/04/A/ST2/00105).
Early stages of this research have also been supported by the
Deutsche Forschungsmeinschaft (DFG, contract Je285/3--2).

\end{document}